\documentclass[conference]{IEEEtran}
\IEEEoverridecommandlockouts

\usepackage{cite}
\usepackage{amsmath}
\usepackage{amssymb}
\usepackage{amsfonts}
\usepackage{algorithmic}
\usepackage{graphicx}
\usepackage{textcomp}
\usepackage[dvipsnames]{xcolor}
\usepackage{hyperref}
\usepackage{arydshln}
\usepackage{booktabs}
\usepackage{cleveref}
\usepackage[labelfont=bf, labelsep=colon]{caption}

\AtEndPreamble{
    \crefname{figure}{Fig.}{Figs.}
    \Crefname{figure}{Fig.}{Figs.}
    \crefname{table}{TABLE}{TABLES}
    \Crefname{table}{TABLE}{TABLES}
}

\newcommand{\eg}{\textit{e}.\textit{g}.}
\newcommand{\ie}{\textit{i}.\textit{e}.}
\definecolor{custompink}{rgb}{0.99, 0.03, 0.51}
\def\BibTeX{{
    \rm B\kern-.05em{\sc i\kern-.025em b}\kern-.08em
    T\kern-.1667em\lower.7ex\hbox{E}\kern-.125emX
}}

\usepackage{amsmath,amsfonts,bm}

\def\eqref#1{equation~\ref{#1}}

\def\1{\bm{1}}

\def\vmu{{\bm{\mu}}}

\def\vepsilon{{\bm{\epsilon}}}

\def\vp{{\bm{p}}}

\def\vs{{\bm{s}}}
\def\vt{{\bm{t}}}

\def\mI{{\bm{I}}}

\def\mX{{\bm{X}}}

\DeclareMathAlphabet{\mathsfit}{\encodingdefault}{\sfdefault}{m}{sl}
\SetMathAlphabet{\mathsfit}{bold}{\encodingdefault}{\sfdefault}{bx}{n}

\DeclareMathOperator*{\argmax}{arg\,max}

\begin{document}

\title{
    Stable-TTS: Stable Speaker-Adaptive Text-to-Speech Synthesis via Prosody Prompting
}

\author{
    \IEEEauthorblockN{
        \textbf{Wooseok Han}$^{*}$\thanks{$^*$Equal contribution}
    }
    \IEEEauthorblockA{
        AITRICS \\ \texttt{hwrg@aitrics.com}
    }
    \and    
    \IEEEauthorblockN{
        \textbf{Minki Kang}$^{*}$
    }
    \IEEEauthorblockA{
        KAIST \\
        \texttt{zzxc1133@kaist.ac.kr}
    }
    \and    
    \IEEEauthorblockN{
        \textbf{Changhun Kim}
    }
    \IEEEauthorblockA{
        KAIST, AITRICS \\
        \texttt{changhun.kim@kaist.ac.kr}
    }
    \and
    \IEEEauthorblockN{
        \textbf{Eunho Yang}
    }
    \IEEEauthorblockA{
        KAIST, AITRICS \\ \texttt{eunhoy@kaist.ac.kr}
    }
}

\maketitle

\begin{abstract}
Speaker-adaptive Text-to-Speech (TTS) synthesis has attracted considerable attention due to its broad range of applications, such as personalized voice assistant services. While several approaches have been proposed, they often exhibit high sensitivity to either the quantity or the quality of target speech samples. To address these limitations, we introduce Stable-TTS, a novel speaker-adaptive TTS framework that leverages a small subset of a high-quality pre-training dataset, referred to as prior samples. Specifically, Stable-TTS achieves prosody consistency by leveraging the high-quality prosody of prior samples, while effectively capturing the timbre of the target speaker.
Additionally, it employs a prior-preservation loss during fine-tuning to maintain the synthesis ability for prior samples to prevent overfitting on target samples.
Extensive experiments demonstrate the effectiveness of Stable-TTS even under limited amounts of and noisy target speech samples.\footnote{Speech samples are available at \href{https://Stable-TTS.github.io}{https://Stable-TTS.github.io}.}
\end{abstract}


\section{Introduction}
Recent advancements in zero-shot TTS models have showcased their ability to generate near-human quality speech resembling the voice of any speaker, using only a few seconds of target speech~\cite{NaturalSpeech2,Vall-E,Grad-StyleSpeech,Mega-TTS,YourTTS}. However, these models frequently face two major issues.
First, extensive pre-training on large datasets, comprising thousands of hours of speech, is necessary to achieve high-quality zero-shot TTS for any speaker~\cite{Vall-E}.
Second, despite extensive pre-training on vast amounts of speech corpora, it remains challenging to generate natural-sounding speech that perfectly mirrors the voice of any speaker, particularly given the distribution shift between source corpora and the target speaker.
\begin{figure}[!ht]
    \centering
    \includegraphics[width=.65\linewidth]{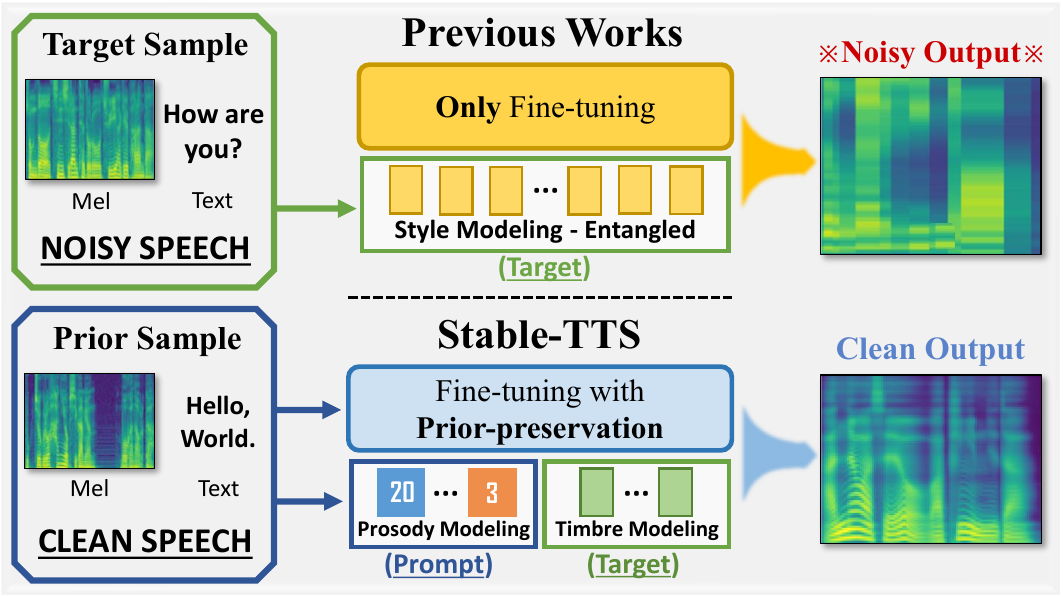}
    \caption{
        \textbf{Concept.}
        Our objective is to build a speaker-adaptive TTS model that utilizes prosody prompt and prior-preservation with prior samples (blue box) to generate a high-quality voice even when fine-tuning with noisy target samples (green box).
    }
    \label{fig:concept}
    \vspace{-0.18in}
\end{figure}

To tackle these challenges, an alternative approach called \emph{transfer learning}, which leverages knowledge from the source domain and involves fine-tuning on speech samples of target speaker, has been widely employed in TTS synthesis~\cite{UnitSpeech,AdaSpeech,AdaptingTTS}.
These models have achieved high stability with small pre-training datasets and effective personalization through fine-tuning on diverse few-shot speech samples, usually under a minute.
However, they often struggle with generating speech that satisfies both \emph{naturalness} and \emph{speaker similarity} criteria, particularly in situations where the few-shot samples suffer from issues such as noise, clipping, or distortion.
For instance, fine-tuning a TTS model with `in-the-wild' recordings~\cite{VoxCeleb} is challenging due to unclear pronunciation and severe background noise, unlike the clear speech data in pre-training~\cite{LibriTTS}. In such cases, fewer fine-tuning steps may reduce voice similarity,
while excessive fine-tuning can compromise the TTS model's integrity to produce clear and intelligible speech.

\begin{figure*}
    \centering
    \includegraphics[width=0.95\linewidth]{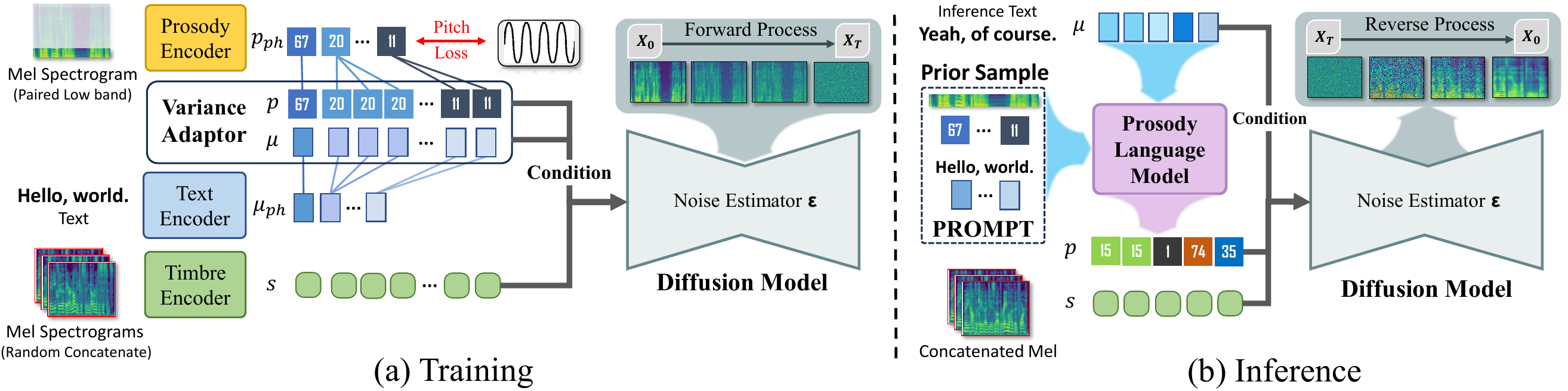}
    \caption{\textbf{Overview of Stable-TTS.} (a) During training, we utilize both the prosody encoder and timbre encoder to enhance timbre and ensure prosody consistency. All representations are utilized as a condition for the diffusion model. (b) During the inference phase, we leverage a Prosody Language Model (PLM) to predict the prosody code utilizing the prompt from a prior sample and generate a timbre vector by concatenating multiple mel-spectrograms derived from the same speaker.}
    \vspace{-0.2in}
    \label{fig:main_fig}
\end{figure*}

Motivated by these limitations, we introduce \emph{Stable-TTS}, a novel speaker-adaptive TTS framework designed to ensure prosody consistency and enhance timbre, even under conditions of limited amount of noisy target samples. Stable-TTS achieves this by leveraging a small subset of high-quality \emph{prior samples} from the pre-training dataset. Specifically, Stable-TTS utilizes a prosody encoder and a Prosody Language Model (PLM)~\cite{Mega-TTS} to guide prosody generation while using a style encoder primarily as a timbre encoder to reinforce the target speaker's timbre. Furthermore, we exploit a \emph{prior-preservation loss}~\cite{Dreambooth} during fine-tuning to maintain synthesis ability for prior samples, thereby preventing overfitting on target samples.
Extensive experiments demonstrate the effectiveness of Stable-TTS in producing high-quality speech samples, excelling in intelligibility, naturalness, and speaker similarity, even under challenging conditions of limited and noisy target samples.

To summarize, our contributions are as follows:
\begin{itemize}
    \item We propose Stable-TTS, a speaker-adaptive TTS framework using a small subset of the high-quality prior samples from a pre-training dataset.

    \item Stable-TTS leverages prior samples by incorporating prosody encoder and PLM to ensure prosody consistency, and employs prior-preservation loss to prevent overfitting.
    
    \item Extensive experiments show that Stable-TTS produces high-quality speech with strong naturalness and speaker similarity, even with limited and noisy target samples.
\end{itemize}

\section{Stable-TTS}
In this section, we introduce Stable-TTS, a speaker-adaptive TTS framework that ensures prosody consistency even with limited noisy target speech samples.
This is achieved by leveraging high-quality \emph{prior samples} from the pre-training dataset during both fine-tuning and inference stages.
The architecture of Stable-TTS incorporates prior samples to stabilize the fine-tuning process, as shown in \Cref{fig:main_fig}, allowing for high-quality, consistent output even under challenging conditions.
The following sections detail the architecture (\Cref{sec:tts}), the use of prior samples in inference (\Cref{sec:ppp}), and fine-tuning stabilization (\Cref{sec:ppl}).

\subsection{Diffusion-Based Zero-Shot TTS Models}
\label{sec:tts}
Our model stems from diffusion-based zero-shot TTS models~\cite{Grad-StyleSpeech, GradTTS}.
Specifically, our model consists of five key modules: text encoder $T$, prosody encoder $P$~\cite{Mega-TTS}, variance adaptor $V$~\cite{FastSpeech, Align}, timbre encoder $S$~\cite{StyleSpeech}, and the diffusion model~\cite{GradTTS}.
To train this model, we use the training dataset consisting of paired samples of the text $\vt$ in the form of phoneme sequences and its speech $\mX$ in the form of a mel-spectrogram.
First, the text encoder $T$ and prosody encoder $P$ encode text $\vt$ and speech $\mX$ into latent representations, $\vmu_{ph} = T(\vt)$ and $\vp_{ph} = P(\mX)$, respectively. Here, $\vp_{ph}$ represents prosody codes at the phoneme level, matching the length of $\vmu_{ph}$. 
Notably, the prosody encoder includes a vector quantization layer~\cite{VQ} at the end, to discretize the prosody representations into the fixed set of discrete prosody codes.
The variance adaptor $V$ then expands both $\vmu_{ph}$ and $\vp_{ph}$ to sequences $\vmu =  V(\vmu_{ph})$ and $\vp = V(\vp_{ph})$ respectively, where both sequences are in the same length with the target mel-spectrogram $\mX$.

Next, the timbre encoder $S$ extracts timbre features from mel-spectrograms.
During training, we sample $k$ mel-spectrograms from the same speaker as $\mX$ and concatenate them to strengthen the timbre characteristics of the target speaker~\cite{Mega-TTS} using the timbre encoder
as follows:
\begin{equation}
\label{eqn:timbre}
    \vs = S(\mX_{R}), \quad \mX_{R} = \texttt{concat}(\mX_{r_1}, \ldots, \mX_{r_k}),
\end{equation}
where $\texttt{concat}$ is the concatenation operator and $\mX_{r_1}$, $\ldots$ , $\mX_{r_k}$ are the mel-spectrograms from the target speaker.

The diffusion model~\cite{DDPM, SDE} generates the speech, using a noise estimator $\vepsilon$ trained to estimate noise during the forward process from the noisy mel-spectrogram.
The forward process is modeled by the following stochastic differential equation, which gradually transforms mel-spectrogram $\mX$ into $\mX_T$:
\vspace{-0.05in}
\begin{equation}
    d\mX_t = -\frac{1}{2}\mX_t \beta_t dt + \sqrt{\beta_t} dW_t, \quad t \in [0, T],
\end{equation}
where $\mX_T$ is a random noise drawn from Gaussian distribution $\mathcal{N}(0, \mI)$, $\beta_t$ is a noise scheduling constant, and $W_t$ is the standard Wiener process for randomness.
In a certain timestep $t$, the noise estimator $\vepsilon$ is trained to estimate the noise given the representations as follows:
\begin{equation}
\label{eqn:diffusion}
    \mathcal{L}_{diff} = \mathbb{E}_{t, \mX, \epsilon} \Vert \vepsilon (\mX_t, t, \vmu, \vp, \vs) + \epsilon\sigma_t^{-1}  \Vert_2^2,
\end{equation}
where $\mX_t$ is a forwarded noisy input $\mX$ at timestep $t$, $\sigma_t = \sqrt{1 - e^{-\int_0^t \beta_s ds}}$, and $\epsilon$ is a noise sampled from $\mathcal{N}(0, \sigma_t^2)$.

\vspace{-0.05in}
\subsection{Prosody Language Model for Prior Prosody Prompting}
\label{sec:ppp}
\vspace{-0.02in}

As detailed in~\Cref{sec:tts}, our TTS model encodes prosody $\vp$ and timbre $\vs$ into distinct representations\cite{Mega-TTS, ProsoSpeech}. 
The timbre vector $\vs$ is continuous and unconstrained, while the prosody codes $\vp$ are discrete and limited to a predefined codebook.
These codes are designed to align with the input text $\vt$, as their purpose is to apply appropriate prosodic elements to the textual content.

During inference, since the target speech and input text are not directly aligned, an additional module is required to predict the prosody codes for the input text.
Mega-TTS~\cite{Mega-TTS} introduced a Prosody Language Model (PLM) to predict prosody codes in an auto-regressive manner, similar to a decoder-only language model used in GPT-2~\cite{GPT2}. The PLM is trained with a language modeling objective~\cite{Mega-TTS}, predicting the prosody codes $\vp$, given the prompt speech $\tilde{\mX}$, as follows: 
\begin{equation}
\label{eqn:plm}
    \argmax_{\vp_{t}} \texttt{PLM}(\vp_{t} | \tilde{\vp}, \tilde{\vmu}, \vp_{<t}, \vmu_{\le t}),
\end{equation}
where $\texttt{PLM}$ is the trained PLM, $\tilde{\vp} = V(P(\tilde{\mX}))$ represents the prompt prosody codes derived from the prompt speech $\tilde{\mX}$, and $\tilde{\vmu} = V(T(\tilde{\vt}))$ is the encoded text from the prompt speech $\tilde{\vt}$.

By constraining the predicted prosody codes $\vp$ to the predefined set in the codebook, we ensure that the prosody of the synthesized speech remains consistent with the pre-training samples, even after fine-tuning.
Moreover, instead of using target speech as the prompt speech $\tilde{\mX}$~\cite{Mega-TTS}, we propose leveraging prior samples---clean speech from the pre-training dataset as a prompt. This allows the PLM to generate a more stable sequence of prosody codes by leveraging those encountered during training. This approach is particularly effective when the target speech is noisy or of poor quality.

\vspace{-0.05in}
\subsection{Prior-Preservation Loss for Fine-Tuning}
\label{sec:ppl}
\vspace{-0.02in}
Fine-tuning the TTS model with the target speaker’s speech samples is crucial for accurately cloning the speaker's voice while preserving naturalness~\cite{AdaSpeech}. However, this becomes challenging when the available samples are \textit{limited} and \textit{noisy}, which is common in real-world recordings.
A well-known issue is that fine-tuning on a few noisy samples can lead to overfitting, causing the model to generate noisy speech~\cite{Grad-StyleSpeech}.

To address this, we focus on fine-tuning the diffusion model, considering its critical role in synthesizing speech from representations.
However, overfitting to limited noisy samples often results in noisy outputs and reduces the effectiveness of prosody codes, making it harder to generate clean speech.
\begin{figure}[t]
    \centering
    \includegraphics[width=0.83\linewidth]{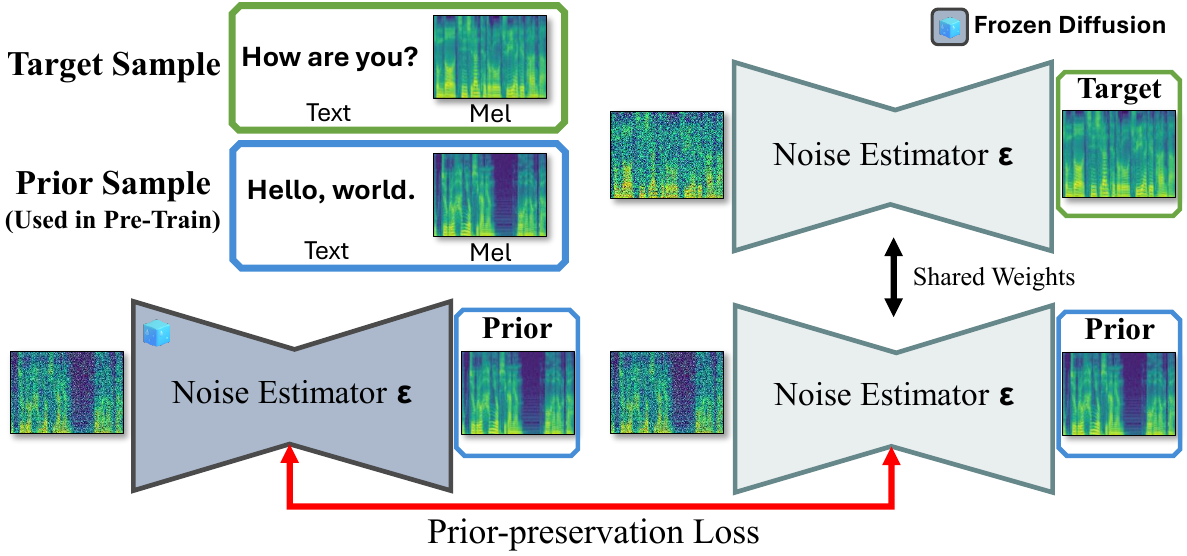}
    \caption{\textbf{Details of Fine-tuning.} We train the diffusion models only using both diffusion and prior-preservation loss.}
    \vspace{-0.2in}
    \label{fig:fine-tuning}
\end{figure}

To mitigate this, we introduce \textit{prior-preservation loss} during fine-tuning, inspired by a personalized text-to-image diffusion model~\cite{Dreambooth}.
This loss maintains the pre-trained distribution by minimizing the difference in estimated noise between clean and noisy samples. As illustrated in~\Cref{fig:fine-tuning}, we use a subset of high-quality \emph{prior samples} from the pre-training dataset, aiming to minimize the mean squared error between the estimated noise of two diffusion models---one with frozen pre-trained weights and the other with modified weights during fine-tuning---based on these prior samples as follows:
\begin{equation}
    \mathcal{L}_{ppl} = \mathbb{E}_{t, \mX} {\Vert \vepsilon(\mX_t, t, \vmu, \vp, \vs; \theta') - \vepsilon (\mX_t, t, \vmu, \vp, \vs; \theta) \Vert}_2^2,
\end{equation}
where $\mX$ is prior samples from the pre-training dataset, and $\mX_t$ is a forwarded noisy input $\mX$ at timestep $t$. In this process, $\theta'$ is the pre-trained and frozen parameter of the noise estimator, and $\theta$ is a parameter to be fine-tuned. The prior-preservation loss ensures that the diffusion model retains the ability to synthesize clean speech. 
During fine-tuning, the noise estimator is fine-tuned by minimizing both $\mathcal{L}_{diff}$ in~\Cref{eqn:diffusion} and an auxiliary loss $\mathcal{L}_{ppl}$.
This approach allows the diffusion model to generate high-quality speech with the voice of the target speaker, even when only a few noisy samples of the target speaker are available.

\section{Experiments} \label{sec:exp}

\begin{table*}
\centering
\caption{
    \textbf{Subjective and Objective Experimental Results} on the \textbf{fine-tuned} models on the LibriTTS test-clean~\cite{LibriTTS}, VCTK~\cite{VCTK}, and VoxCeleb~\cite{VoxCeleb} speech dataset of less than one minute for each speaker. Speech samples used in this human evaluation are randomly selected. We report the training dataset for a fair comparison and note that UnitSpeech$^\dagger$~\cite{UnitSpeech} is trained on the LibriTTS train set~\cite{LibriTTS}. Evaluation for ground truth on VoxCeleb is unavailable since all but a limited number of target samples are used.
}
\resizebox{\textwidth}{!}{
\begin{tabular}{lcccccccccccc}
    \toprule
    
    & \multicolumn{4}{c}{\textbf{LibriTTS test-clean}} & \multicolumn{4}{c}{\textbf{VCTK}} & \multicolumn{4}{c}{\textbf{VoxCeleb}}  \\
    
    \cmidrule(lr){2-5} \cmidrule(lr){6-9} \cmidrule(lr){10-13}
    
    {\textbf{Model}}  &  MOS ($\uparrow$) & SMOS ($\uparrow$) &  WER ($\downarrow$)  &  SECS ($\uparrow$) &  
                         MOS ($\uparrow$) & SMOS ($\uparrow$) &  WER ($\downarrow$)  &  SECS ($\uparrow$) &  
                         MOS ($\uparrow$) & SMOS ($\uparrow$) &  WER ($\downarrow$)  &  SECS ($\uparrow$) \\
    
    \midrule[0.8pt]
    
    Ground Truth & 3.99$_{\pm \text{0.20}}$ & 3.61$_{\pm \text{0.21}}$ &   4.76$_{\pm \text{1.09}}$  &  83.08$_{\pm \text{0.08}}$ & 
                    4.17$_{\pm \text{0.17}}$ & 3.95$_{\pm \text{0.21}}$ &   2.96$_{\pm \text{0.90}}$  &  90.90$_{\pm \text{0.10}} $  & 
                     - & -                                              &   
                     - & - \\
    
    \midrule
    
    Grad-StyleSpeech~\cite{Grad-StyleSpeech} & 3.11$_{\pm \text{0.22}}$ & 3.19$_{\pm \text{0.23}}$  &   11.31$_{\pm \text{1.52}}$  &  85.69$_{\pm \text{0.69}}$  & 
                                               2.88$_{\pm \text{0.20}}$ & 2.94$_{\pm \text{0.22}}$  &   14.55$_{\pm \text{2.01}}$  &  \bf 87.19$_{\pm \text{0.19}}$  & 
                                               2.60$_{\pm \text{0.21}}$ & 2.86$_{\pm \text{0.22}}$  &   16.06$_{\pm \text{1.98}}$  &  \bf 85.27$_{\pm \text{0.27}}$ \\
    
    UnitSpeech$^\dagger$~\cite{UnitSpeech} & 2.16$_{\pm \text{0.20}}$ & 2.55$_{\pm \text{0.22}}$  &   6.00$_{\pm \text{1.92}}$   &  80.39$_{\pm \text{1.39}}$  & 
                                               2.49$_{\pm \text{0.21}}$ & 2.94$_{\pm \text{0.22}}$  &   4.64$_{\pm \text{1.11}}$ &  80.89$_{\pm \text{0.89}}$  & 
                                               2.63$_{\pm \text{0.21}}$ & 2.59$_{\pm \text{0.24}}$  &   19.76$_{\pm \text{5.08}}$  &  73.81$_{\pm \text{1.81}}$ \\
    
    \midrule
    
    \textbf{Stable-TTS} (w/o PLM)   & 3.16$_{\pm \text{0.23}}$ & 3.21$_{\pm \text{0.26}}$      &  11.50$_{\pm \text{1.51}}$     &  \bf 85.75$_{\pm \text{0.75}} $  & 
                                       2.65$_{\pm \text{0.21}}$ & 2.95$_{\pm \text{0.22}}$      &  14.35$_{\pm \text{2.03}}$     &  87.13$_{\pm \text{0.13}}$  & 
                                       2.59$_{\pm \text{0.20}}$ & \bf 2.93$_{\pm \text{0.22}}$  &  14.06$_{\pm \text{1.78}}$     &  84.73$_{\pm \text{0.73}}$ \\
    
    \textbf{Stable-TTS} (w/o P.P.)   & 2.73$_{\pm \text{0.24}}$ & 2.73$_{\pm \text{0.24}}$      &  3.16$_{\pm \text{1.04}}$      &  82.16$_{\pm \text{1.16}}$  & 
                                       3.30$_{\pm \text{0.20}}$ & 3.55$_{\pm \text{0.20}}$      &  1.69$_{\pm \text{0.64}}$      &  85.36$_{\pm \text{0.36}}$  & 
                                       2.69$_{\pm \text{0.23}}$ & 2.53$_{\pm \text{0.22}}$      &  5.64$_{\pm \text{1.45}}$      &  81.85$_{\pm \text{0.85}}$ \\

    \textbf{Stable-TTS} (w/o Prior Prompt) & \bf 3.65$_{\pm \text{0.21}}$    & 3.37$_{\pm \text{0.21}}$      &  \bf 0.83$_{\pm \text{0.37}}$      &  83.16$_{\pm \text{0.16}}$  & 
                                            3.34$_{\pm \text{0.19}}$    & 3.76$_{\pm \text{0.20}}$      &  0.72$_{\pm \text{0.46}}$      &  85.41$_{\pm \text{0.41}}$  & 
                                            2.62$_{\pm \text{0.23}}$    & 2.44$_{\pm \text{0.22}}$      &  \bf 1.02$_{\pm \text{0.46}}$      &  82.04$_{\pm \text{1.04}}$ \\

    \textbf{Stable-TTS}  &   3.37$_{\pm \text{0.20}}$ & \bf 3.41$_{\pm \text{0.22}}$   &   {1.13}$_{\pm \text{0.42}}$  &  83.13$_{\pm \text{1.13}} $  & 
                               \bf 3.82$_{\pm \text{0.19}}$ & \bf 4.04$_{\pm \text{0.22}}$   &   \bf {0.49}$_{\pm \text{0.34}}$  &  85.26$_{\pm \text{0.26}}$   & 
                               \bf 3.00$_{\pm \text{0.22}}$ & 2.65$_{\pm \text{0.23}}$       &   1.32$_{\pm \text{0.50}}$  &  82.09$_{\pm \text{1.09}}$ \\
    
    \bottomrule
    
        \end{tabular}}
\label{tab:main_table}
\end{table*}

\subsection{Experimental Setup}
\emph{Datasets.}\quad
Stable-TTS is pre-trained on the clean-100 and clean-360 subsets of LibriTTS-R~\cite{LibriTTS-R}, an English multi-speaker dataset with 245 hours of audio from 553 speakers. For fine-tuning, we utilize the configuration from 24 speakers in the test-clean subset of LibriTTS~\cite{LibriTTS}, and the configuration from 24 speakers in the VCTK~\cite{VCTK} dataset, with each containing 20 audio clips averaging 1-4 seconds in length.
Furthermore, to validate Stable-TTS in real-world noisy scenarios, we utilize 20 speakers from VoxCeleb~\cite{VoxCeleb}, comprising audio-visual short clips extracted from the interview video dataset. The transcripts are generated using Whisper~\cite{whisper} ASR, and each dataset contains 5-7 audio clips, ranging from 4-8 seconds in length.

\emph{Baselines.}\quad
We compare Stable-TTS with two recent diffusion-based TTS models, Grad-StyleSpeech~\cite{Grad-StyleSpeech} and UnitSpeech~\cite{UnitSpeech}. \emph{Grad-StyleSpeech} is an any-speaker adaptive TTS model, comprising a diffusion and a style-adaptive encoder with style-adaptive layer normalization~\cite{StyleSpeech}. \emph{UnitSpeech}, based on Grad-TTS~\cite{GradTTS}, enables speaker adaptation with untranscribed speech using self-supervised unit representation as a pseudo transcript, along with a unit encoder. 

\emph{Evaluation metrics.}\quad
For objective evaluation, we employ Word Error Rate (WER) to assess the intelligibility of synthesized speech and Speaker Embedding Cosine Similarity (SECS) to measure the speaker similarity, utilizing speaker verification model of Resemblyzer~\cite{Resemblyzer}.
For subjective evaluation, we employ Mean Opinion Score (MOS) for naturalness and Similarity Mean Opinion Score (SMOS) for speaker similarity, where 20 evaluators rate the synthesized speech on a scale from 1 to 5, measuring each attribute separately.

\emph{Implementation details.}\quad
We set the audio config to 16khz sampling rate and 80 mel bin.
Our diffusion-based zero-shot TTS model, based on Grad-StyleSpeech~\cite{Grad-StyleSpeech}, closely follows its pre-training setup.
For the prosody encoder, we use a vector quantization module, using a low band with rich prosodic contents of size 15 from paired mel-spectrogram as input for the prosody vector. As mentioned in \Cref{sec:tts}, three random mel-spectrograms from the same speaker are concatenated into one for the timbre vector input, \ie, $k=3$ in \Cref{eqn:timbre}.
We randomly choose prior samples from the pre-training dataset, where each sample has a mel-spectrogram length ranging from 100 to 150. 
Moreover, for the prior prosody prompts in inference, we carefully choose one speech sample for each male and female which are exclusively used in most cases.
We adopt HiFi-GAN~\cite{HifiGAN} as a vocoder.

\subsection{Main Results}
We compare the performance of Stable-TTS against baselines across diverse datasets to ascertain its efficacy in enhancing intelligibility, naturalness, and similarity. As shown in \Cref{tab:main_table}, Stable-TTS exhibits superior performance in terms of both MOS and SMOS across all datasets. It is worth noting that Stable-TTS significantly reduces the WER of existing baselines (by 81\%, 89\%, and 92\% for three datasets, respectively), demonstrating exceptional improvement in intelligibility. This confirms that Stable-TTS performs excellently not only with clean speech samples (LibriTTS, VCTK) but also with noisy speech samples (VoxCeleb). Remarkably, this achievement has been accomplished with only a slight (VCTK, VoxCeleb) or no sacrifice (LibriTTS) in SECS, indicating that Stable-TTS significantly outperforms other baselines in terms of the intelligibility-similarity trade-off.

\subsection{Ablation Study}
\emph{Components of Stable-TTS.}\quad
We validate the core strategies of Stable-TTS, namely PLM for prior prosody prompting (\Cref{sec:ppp}) and prior-preservation loss (P.P., \Cref{sec:ppl}) as well as utilizing prior samples (\Cref{sec:ppl}). As shown in \Cref{tab:main_table}, we find that removing either component results in a degradation of performance in terms of MOS, SMOS, and WER across all scenarios. Notably, even Stable-TTS without P.P. already outperforms baselines, underscoring the efficacy of prosody prompting with PLM.
Removing P.P. maintains SECS but significantly degrades MOS, SMOS, and WER, showing that prior preservation prevents overfitting to the target samples, enabling intelligible and natural speech synthesis.
It is worth highlighting that using prior samples on PLM for prosody prompting rather than target samples significantly enhances naturalness, especially for out-of-domain datasets (VCTK, VoxCeleb).

\begin{table}[ht]
\centering
\footnotesize
\vspace{-0.05in}
\caption{\textbf{Zero-shot vs. Fine-tuning.} Analysis in zero-shot and fine-tuning utilizing 24 speakers from the VCTK~\cite{VCTK}.}
\resizebox{0.55\linewidth}{!}{
\begin{tabular}{lcc}
    \toprule
    
    & \multicolumn{2}{c}{\textbf{Stable-TTS}}  \\
    \cmidrule(lr){2-3}
    
    \bf Setting   & WER ($\downarrow$) & SECS ($\uparrow$) \\
    
    \midrule[0.4pt]
    
    Zero-shot & 0.91$_{\pm \text{0.33}}$  & 84.18$_{\pm \text{0.18}}$ \\

    Fine-tuning & \bf 0.49$_{\pm \text{0.34}}$  &  \bf {85.26$_{\pm \text{0.26}}$} \\
    
    \bottomrule
\end{tabular}
}
\vspace{-0.05in}
\label{tab:vs_zeroshot}
\end{table}

\begin{figure}[ht]
\centering
\vspace{-0.12in}
\includegraphics[width=.83\linewidth]{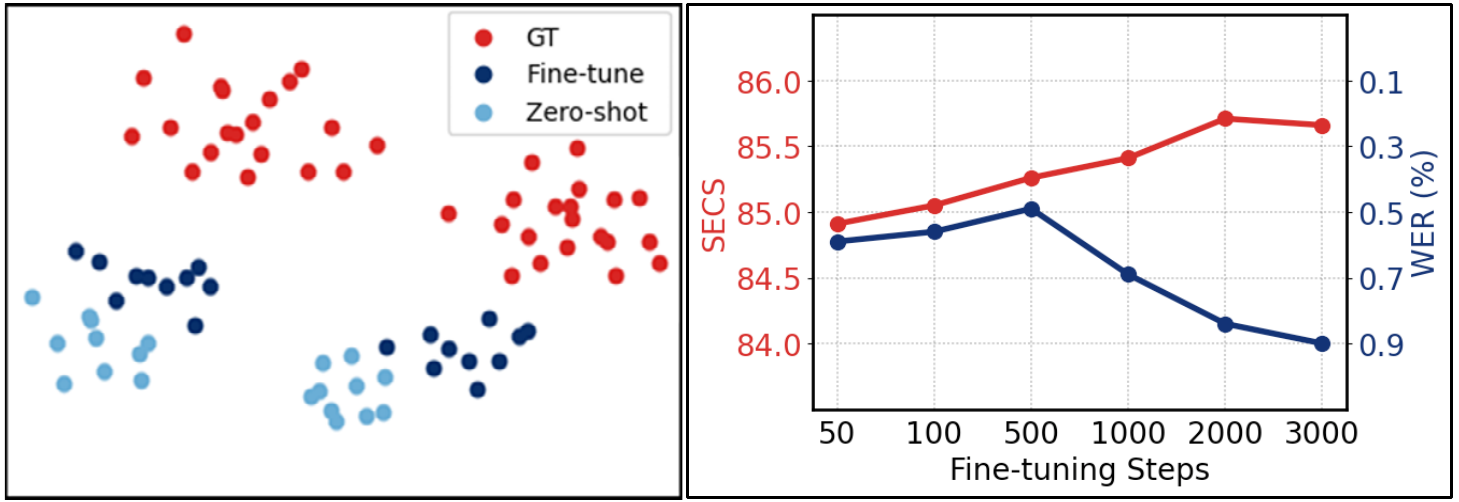}
\vspace{-0.05in}
\caption{\textbf{Zero-shot vs. Fine-tuning.} t-SNE visualization (left) analysis of SECS and WER with fine-tuning iterations (right).}
\label{fig:tsne_ftsteps}
\vspace{-0.17in}
\end{figure}

\emph{Zero-shot vs. fine-tuning.}\quad
Stable-TTS can also be used in a zero-shot setting without fine-tuning on the target speaker. To evaluate fine-tuning's effectiveness, we compare zero-shot performance with fine-tuning improvements.
In \Cref{tab:vs_zeroshot} and \Cref{fig:tsne_ftsteps} (left), Stable-TTS already exhibits commendable performance in zero-shot scenarios. However, fine-tuning enhances both naturalness and similarity.
Further analysis of fine-tuning steps in \Cref{fig:tsne_ftsteps} (right) reveals that fine-tuning is not sensitive to the number of steps, indicating the presence of a `sweet spot'. Specifically, 500 steps appear to strike the optimal balance.

\subsection{Evaluation under Limited Amounts of Target Samples}
Since our objective is to excel not only in noisy datasets (\eg, VoxCeleb in \Cref{tab:main_table}) but also in scenarios with limited amounts of data, we conduct experiments on restricted samples of the target speaker by decreasing the number of samples used during fine-tuning using VCTK dataset. In \Cref{tab:finetune_scale}, we prove that Stable-TTS maintains stable WER of less than 1\% even with limited target samples. This is in stark contrast to Grad-StyleSpeech, where the performance deteriorates as the number of target samples decreases.

\begin{table}[h]
\centering
\vspace{-0.05in}
\footnotesize
\caption{\textbf{Fine-tuning Data Scale.} The experiment involves 10 speakers from the VCTK~\cite{VCTK}, each with more than 100 samples, and the steps are kept consistent.}
\resizebox{0.75\linewidth}{!}{
\begin{tabular}{lcccc}
    \toprule
    
    & \multicolumn{2}{c}{\textbf{Grad-StyleSpeech}~\cite{Grad-StyleSpeech}} & \multicolumn{2}{c}{\textbf{Stable-TTS}} \\
    
    \cmidrule(lr){2-3} \cmidrule(lr){4-5}
    
    \bf \# Samples & WER ($\downarrow$) & SECS ($\uparrow$) & WER ($\downarrow$) & SECS ($\uparrow$) \\
    
    \midrule[0.4pt]
    
    1 & 48.92$_{\pm \text{9.11}}$  &  84.23$_{\pm \text{0.23}}$  & 0.97$_{\pm \text{0.67}}$  & 83.88$_{\pm \text{0.88}}$ \\
    
    5 & 18.66$_{\pm \text{3.40}}$  &  86.87$_{\pm \text{0.87}}$  & 0.52$_{\pm \text{0.53}}$  & \bf {84.97$_{\pm \text{0.03}}$} \\
    
    20 & 18.58$_{\pm \text{3.33}}$  &  \bf 86.93$_{\pm \text{0.07}}$  & 0.35$_{\pm \text{0.42}}$  & 84.74$_{\pm \text{0.74}}$  \\
    
    100 & \bf 14.16$_{\pm \text{3.02}}$  &  86.34$_{\pm \text{0.34}}$  & \bf 0.25$_{\pm \text{0.29}}$  &  84.61$_{\pm \text{0.61}}$ \\
    
    \bottomrule
\end{tabular}
}
\vspace{-0.13in}
\label{tab:finetune_scale}
\end{table}

\section{Conclusion}
In this paper, we have introduced Stable-TTS, a speaker-adaptive TTS framework under limited and noisy speech samples.
The core idea of Stable-TTS is to use a small subset of high-quality \emph{prior samples} from a pre-training dataset. Stable-TTS integrates a prosody language model into our TTS system to conduct prior prosody prompting and incorporates a prior-preservation loss during fine-tuning.
Extensive experiments demonstrated the effectiveness of Stable-TTS in terms of naturalness and speaker similarity, even with limited amounts and poor-quality speech samples. 
\bibliographystyle{IEEEtran}
\bibliography{reference}

\end{document}